\documentclass{mem}
\usepackage{natbib}\usepackage{txfonts}
\usepackage{balance}
\usepackage{graphicx}
\usepackage[a4paper]{hyperref}
\idline{80}{735}
\begin{document}
\def\teff{$T\rm_{eff }$}
\def\kms{$\mathrm {km s}^{-1}$}

\title{
3D molecular line formation in dwarf carbon-enhanced metal-poor stars
}

   \subtitle{}

\author{
N.~T.~Behara\inst{1,2}, 
H.-G.~Ludwig\inst{1,2},
P.~Bonifacio\inst{1,2,3},
L.~Sbordone\inst{1,2},
J.~I.~Gonz\'alez~Hern\'andez\inst{1,2},
\and E.~Caffau\inst{2}
          }

  \offprints{N.~T.~Behara} 

\institute{
CIFIST Marie Curie Excellence Team
\and
GEPI, Observatoire de Paris
CNRS, Universit\'e Paris Diderot\\ 
Place Jules Janssen 92190
Meudon, France 
\and
Istituto Nazionale di Astrofisica - Osservatorio Astronomico di Trieste\\
Via Tiepolo 11, I-34143  Trieste, Italy\\
\email{natalie.behara@obspm.fr}
}

\authorrunning{Behara}

\titlerunning{3D molecular lines in CEMP stars}

\abstract{
We present a detailed analysis of the carbon and nitrogen abundances of two
dwarf carbon-enhanced metal-poor (CEMP) stars: SDSS J1349-0229 and SDSS
J0912+0216. We also report the oxygen abundance of SDSS J1349-0229. These stars are
metal-poor, with [Fe/H] $<$ --2.5, and were selected from our
ongoing survey of extremely metal-poor dwarf candidates from the Sloan Digital Sky
Survey (SDSS). 
The carbon, nitrogen and oxygen abundances rely on molecular lines
which form in the outer layers of the stellar atmosphere. It is known that
convection in metal-poor stars induces very low temperatures which are not predicted by
`classical' 1D stellar atmospheres. To obtain the correct temperature
structure, one needs full 3D hydrodynamical models. 
Using CO5BOLD 3D hydrodynamical model atmospheres and the Linfor3D line
formation code, molecular lines of CH, NH, OH  and C$_2$ were computed, and 3D
carbon, nitrogen and oxygen abundances were determined. The resulting
carbon abundances were compared to abundances derived using atomic C {\sc i}
lines in 1D LTE and NLTE. For one star,  SDSS J1349-0229, we were able to compare the
3D oxygen abundance from OH lines to O {\sc i} lines in 1D LTE and NLTE.
There is not a good agreement between the carbon 
abundances determined from C$_2$ bands and from the CH band, 
and
molecular lines do not agree with the atomic C {\sc i} lines.
Although this may be partly due to 
uncertainties in the transition probabilities of the
molecular bands it certainly has to do with the temperature
structure of the outer layers of the adopted model atmosphere.
In fact the discrepancy between C$_2$ and CH is in opposite
directions when using 3D and 1D models.
Confronted with this inconsistency, we explore the influence of the 3D
model properties on the molecular abundance determination. In particular, the
choice of the number of opacity bins used in the model calculations and its
subsequent effects on the temperature structure and molecular line formation
is discussed.  

}
\maketitle{}

\section{Introduction}

SDSS J1349-0229 and SDSS J0912+0216 were selected from our ongoing survey of
extremely metal-poor star candidates from the Sloan Digital Sky
Survey. 
SDSS J1349-0229 has an effective temperature of 6200~K, log $g$ = 4.00 and
[Fe/H] = --3.0. SDSS J0912+0216 is slightly hotter with a temperature
of 6500~K, log $g$ = 4.50 and [Fe/H] = --2.5. 
These two stars belong to the class CEMP-r+s, since their
atmospheres are enhanced in carbon, and show
overabundances of $s$ and $r$-process elements. For details on 
the abundances of the neutron-capture elements, see Behara et
al.~(2009).

The focus of this study is on the abundances of CEMP stars determined
predominately using molecular lines: carbon, nitrogen and oxygen. It
is known that overcooling in the outermost layers of metal-poor stars
produced by convection is not predicted by classical 1D stellar
atmospheres and is best reproduced by 3D hydrodynamical
simulations (Asplund et al.~1999; Caffau \& Ludwig 2007; Gonz\'alez
Hern\'andez et al.~2008). An accurate description of these outer
layers is crucial for the determination of the molecular abundances.

Fortunately these two CEMP stars are hot enough to display
atomic carbon lines. These lines are formed much deeper in the
atmosphere compared to molecular lines, and are thus shielded from
the overcooled region of the atmosphere, allowing for a validity check
of the abundances obtained from molecular lines.

\section{Abundance analysis}

In our analysis, we used 3D model atmospheres computed with the
CO$^5$BOLD code (Freytag et al.~2002; Wedemeyer et al.~2004). The 3D
spectral synthesis calculations were performed with the code
Linfor3D. The parameters of the 3D models used are listed in Table~\ref{tab1}. 

 \begin{table}
 \caption{ 
 Parameters of the 3D models. The model $a$ was
 used in the analysis of SDSS J1349-0229. We interpolated between the
 corrections from the models $b$ for
 SDSS J0912+0216. The first five models were computed using 6 opacity bins, while
 the last three using 12 bins - see discussion for details.}
 \label{tab1}
 \begin{center}
 \begin{tabular}{lccc}
 Model        & T$_{\rm eff}$ & log $g$ & [Fe/H] \\
 \hline 
 d3t63g40mm30n01$^a$ & 6270 & 4.00 & --3.0 \\
 d3t65g45mm20n01$^b$ & 6530 & 4.50 & --2.0 \\
 d3t65g45mm30n01$^b$ & 6550 & 4.50 & --3.0 \\
 d3t63g40mm20n01 & 6280 & 4.00 & --2.0 \\
 d3t63g40mm10n01 & 6260 & 4.00 & --1.0 \\
 \hline
 d3t63g40mm30n02 & 6240 & 4.00 & --3.0 \\
 d3t63g40mm20n02 & 6250 & 4.00 & --2.0 \\
 d3t63g40mm10n02 & 6250 & 4.00 & --1.0 \\
\hline
 \end{tabular}
 \end{center}
 \end{table}

We compared each of our 3D models to a corresponding standard
hydrostatic 1D model, and define the 3D correction of an
abundance measure in the sense 3D -- 1D. The 1D models are calculated using
plane-parallel geometry and employ the same equation-of-state and
opacities as the CO$^5$BOLD models. 

With the exception of C$_2$, all of the abundances were determined
using equivalent widths of unblended lines. The C$_2$ abundance was
determined by line fitting using the same code used by 
\citet{caffau05}. For CH, we adopted the line list
used by Bonifacio et al.~(1998), for the UV OH lines the $gf$ values have  
been computed from the lifetimes calculated by Goldman \& Gillis (1981), and for
C$_2$ and NH we used the molecular line lists by Kurucz (2005). The results
for the two stars are listed in Table~\ref{tab2}.

 \begin{table}
 \caption{
 3D and 1D abundances expressed with respect to solar abundances of log\,($\epsilon$)\,=\,8.50,
 7.86, and 8.76 for carbon, nitrogen and oxygen, respectively.} 
 \label{tab2}
 \begin{center}
 \begin{tabular}{l cc}
              & \multicolumn{2}{c}{SDSS J1349-0229} \\  
 \hline 
 Element      & [X/Fe]$_{\rm 1D}$ & [X/Fe]$_{\rm 3D}$ \\
 \hline
 CH           & 2.82 & 2.09  \\
 C$_2$        & 3.16 & 1.72  \\
 C {\sc i} {\small NLTE}   & 2.42 & 2.51 \\ 
 NH           & 1.60 & 0.67  \\
 OH           & 1.88 & 1.70  \\
 O {\sc i} {\small NLTE}   & 1.63 & 1.69 \\
 \hline
              & \multicolumn{2}{c}{SDSS J0912+0216}  \\  
 \hline
 CH           & 2.17 & 1.67  \\
 C {\sc i} {\small NLTE}   & 1.38 &  1.44 \\ 
 NH           & 1.75 & 1.07  \\
 \hline
 \end{tabular}
 \end{center}
 \end{table}

\section{Discussion}

\begin{figure}[]
\resizebox{\hsize}{!}{\includegraphics[clip=true]{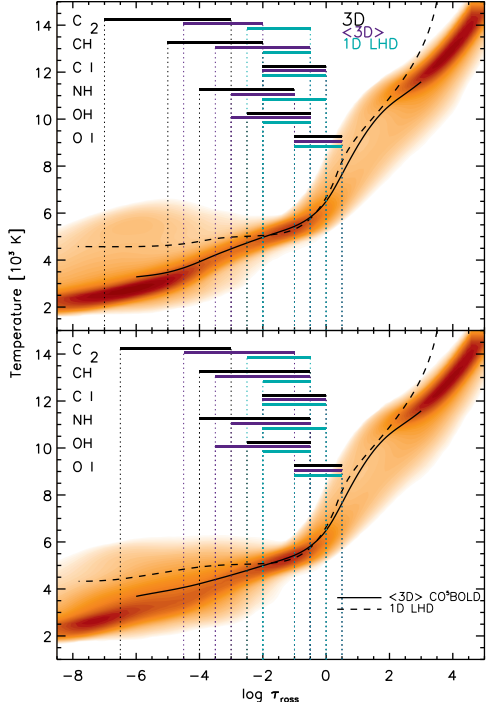}}
\caption{\footnotesize
 The temperature structures of the 3D 6 bin model d3t63g40mm30n01 (top) and 12
 bin model d3t63g40mm30n02 (bottom) are plotted as a function of optical depth
 along with the average 3D (solid line) and corresponding 1D (dashed line) temperature
 structures. Overplotted on the figure are the ranges of depth of formation of
 the C$_2$, CH, C {\sc i}, NH, OH and O {\sc i} spectral lines used in the analysis.}
\label{fig1}
\end{figure}

The sensitivity of the molecular lines to temperature is clear from the
significant 3D corrections. In Figure~\ref{fig1} we have plotted the
temperature distributions for the models used in the analysis of SDSS
J1349-0229. The ranges of depth of formation of 
the lines used in the analysis are overplotted for comparison. 
The different depth of formation of the 
C$_2$ and CH lines in the 1D and 3D models
explains why in 1D we derive a larger C abundance
from the C$_2$ lines than from the CH lines, while in 3D the
reverse is true. The C$_2$ lines  
are formed much higher up in the atmosphere compared to the CH lines,
and this tendency is much larger in the 3D than in the 1D model. 
The 3D models on the other hand do not achieve a better consistency between
CH and C$_2$ lines.
The C {\sc i}
lines are quite insensitive to 3D effects. 
Neither in 3D nor in 1D
we  achieve consistent results between molecular lines
and C {\sc i} lines. 

\balance
To shed light on the discrepancy between the carbon abundance
indicators, we use our best indicator, C {\sc i}, as a reference and explore
the influence of the 3D model properties on the molecular abundance
determination. In particular, we explore the effect of increasing the number
of opacity bins in the opacity groups from 6 bins to 12
bins as described in \citet{ludwigjd10}. The new abundance
obtained for SDSS J1349-0229 are listed in Table~\ref{tab3}.

 \begin{table}
 \caption{Comparison between 3D abundances derived using models with 6 and 12
   opacity bins. }
 \label{tab3}
 \begin{center}
 \begin{tabular}{l|cc}
 Element &  6 bin [X/Fe]$_{\rm 3D}$ & 12 bin [X/Fe]$_{\rm 3D}$ \\
\hline
 CH & 2.09 &  2.22 \\
 C$_2$ & 1.72 & 1.99 \\
 C {\sc i} {\small NLTE} &  2.51 & 2.50 \\
 NH & 0.67  &  0.87 \\
 OH &  1.69 &  1.69 \\
 O {\sc i} {\small NLTE} & 1.69 & 1.68 \\
\hline
 \end{tabular}
 \end{center}
 \end{table}

The temperature distribution of the 12 bin model is shown in Figure~\ref{fig1} below
the 6 bin model. The different binning scheme results in a decrease of the
overcooling, thus lowering the 3D corrections of the molecules formed higher
in the atmosphere.

\begin{figure*}[]
\begin{center}
\resizebox{\hsize}{!}{\includegraphics[clip=true]{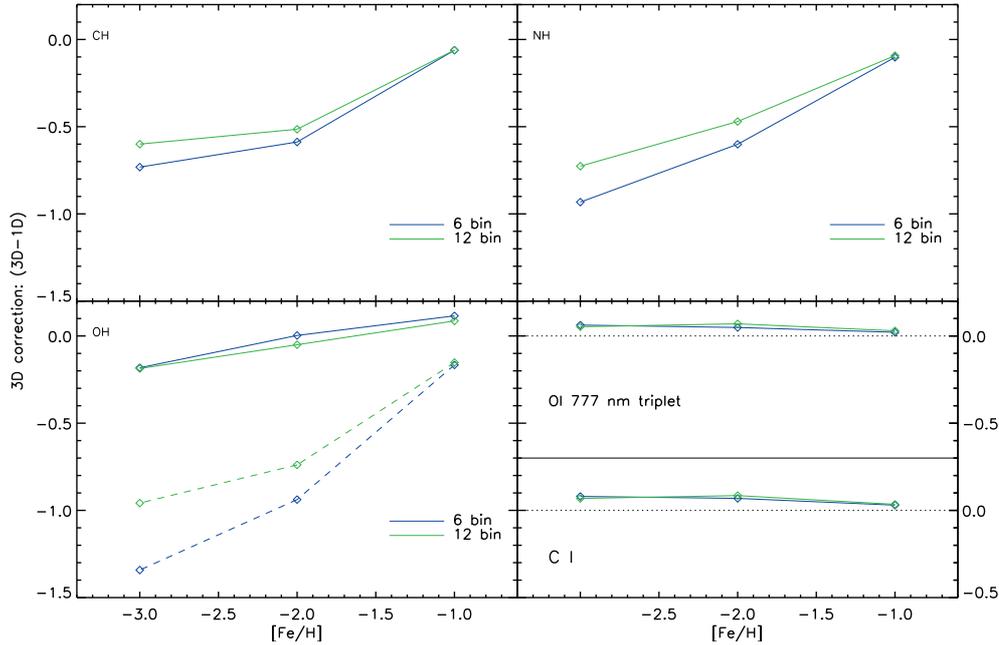}}
\end{center}
\caption{
 3D corrections as a function of metallicity for CH, OH, NH, O {\sc i} and C
 {\sc i}. The 6 bin models are plotted in black, while the 12 bin models are
 plotted in grey. The OH correction have been calculated assuming a typical CEMP
 composition: an enhanced carbon-to-oxygen ratio of +1.0. Overplotted as
 dashed lines are the corrections using scaled solar abundances.}
\label{fig2}
\end{figure*}

The atomic lines, formed deeper in the atmosphere are
unaffected. The 12 bin models give a better agreement between the CH 
the C$_2$ lines and both molecular features
appear to be in better agreement with the C {\sc i}. 
Although these results appear encouraging, the
formation region is so external
and at low density, that the validity of LTE for
molecule formation and levels population should be questioned.

To explore the sensitivity of the models to the different binning schemes, we
compute 3D corrections for CH, C {\sc i}, NH, OH and O {\sc i} for [Fe/H] =
--3.0, --2.0 and --1.0 using the models listed in Table~\ref{tab1}. 
The corrections are shown in Figure~\ref{fig2}. 
The OH lines computed using scaled-solar abundances appear the most
sensitive. In carbon-enhanced atmospheres however, the effect is much
reduced. Overall, with regards to the molecular lines, the sensitivity to the
binning scheme as well as the 3D corrections decrease with increasing
metallicity. The atomic lines proved to be immune to both the 3D model
properties and 3D effects.

\begin{acknowledgements}
We acknowledge financial support from  
EU contract MEXT-CT-2004-014265 (CIFIST).
\end{acknowledgements}

\bibliographystyle{aa}

\end{document}